# Dynamics of episodic supershear in the 2023 M7.8 Kahramanmaraș/Pazarcik earthquake, revealed by near-field records and computational modeling

Mohamed Abdelmeguid[1,6 ✉], Chunhui Zhao[2,6], Esref Yalcinkaya[3], George Gazetas[4], Ahmed Elbanna[2,5 ✉] & Ares Rosakis[1 ✉]

The 2023 M7.8 Kahramanmaraș/Pazarcik earthquake was larger and more destructive than what had been expected. Here we analyzed nearfield seismic records and developed a dynamic rupture model that reconciles different currently conflicting inversion results and reveals spatially non-uniform propagation speeds in this earthquake, with predominantly supershear speeds observed along the Narli fault and at the southwest (SW) end of the East Anatolian Fault (EAF). The model highlights the critical role of geometric complexity and heterogeneous frictional conditions in facilitating continued propagation and influencing rupture speed. We also constrained the conditions that allowed for the rupture to jump from the Narli fault to EAF and to generate the delayed backpropagating rupture towards the SW. Our findings have important implications for understanding earthquake hazards and guiding future response efforts and demonstrate the value of physics based dynamic modeling fused with near-field data in enhancing our understanding of earthquake mechanisms and improving risk assessment.

[1] Graduate Aerospace Laboratories, California Institute of Technology, 1200 E. California Boulevard, Pasadena 91125 CA, USA. [2] Department of Civil and Environmental Engineering, University of Illinois at Urbana Champaign, 205 N. Mathews Avenue, Urbana 61801 IL, USA. [3] Department of Geophysical Engineering, Istanbul University-Cerrahpasa, 34320 Avcilar, Istanbul, Turkey. [4] Department of Civil Engineering, National Technical University of Athens, 9, Iroon Polytechniou Str, Athens, Greece. [5] Beckman Institute of Advanced Science and Technology, University of Illinois at Urbana Champaign, 405 N. Mathews Avenue, Urbana 61801 IL, USA. [6] These authors contributed equally: Mohamed Abdelmeguid, Chunhui Zhao. ✉email: meguid@caltech.edu; elbanna2@illinois.edu; arosakis@caltech.edu

On February 6, 2023, a $M_w$ 7.8 earthquake, currently known as the Kahramanmaraş/Pazarcik earthquake, shook the southeastern parts of Türkiye and northern Syria. Preliminary back projection models based on teleseismic data as well as multiple seismic inversions suggest that the rupture initiated at 1:17:355 coordinated universal time (UTC) on a splay fault (the Narli fault) in the near proximity of the East Anatolian fault[1,2]. The hypocenter location is estimated by USGS to be 37.230°N 37.019°E with a depth of 10 km[1,2]. The rupture then propagated northeast subsequently transferring to the East Anatolian fault and starting a sequence of seismic events. Furthermore, subsequent preliminary geodetic inversions confirmed the multi-segment nature of the $M_w$ 7.8 rupture[3,4]. The sequence of events resulted in catastrophic levels of destruction with substantial humanitarian and financial losses[5].

The M7.8 Kahramanmaraş/Pazarcik earthquake was, by many measures, bigger and more destructive than what had been expected based on historical records in the past several centuries[6–11]. The estimated magnitude of the largest earthquake that occurred on the East Anatolian Fault (EAF) in the last few hundred years is 7.2 which is believed to be either the 1789 Palu (Elazığ) earthquake or the 1872 Amanos earthquake[12,13]. This estimate is smaller than the magnitude of the Kahramanmaraş/Pazarcik earthquake. Furthermore, each of these historic events ruptured a segment of the EAF but none was extended over multiple segments as the recent event.

From a geological point of view, there are several features associated with the fault system that could have contributed to the extent of damage associated with the Mw7.8 Kahramanmaraş/Pazarcik earthquake. Studies of the tectonic setting suggest that the orientation of the EAF with respect to the principal stresses places several fault segments within a highly stressed regime[14]. This stress regime is sensitive to minor perturbations associated with dynamic stress transfer and dynamic stress rotations. Furthermore, the fault network is geometrically complex with multiple fault segmentations, kinks, and bends[15–17] which strongly influences the dynamics of rupture propagation[18–22]. The existence of geometrical complexity within this high-stress regime could further amplify its role in rupture dynamics through, for example, the emergence of regions with high-stress concentrations, generation of arrest phases, backpropagation of earthquake rupture, or development of episodes of transient supershear propagation.

Our preliminary analysis of the Kahramanmaraş/Pazarcik earthquake based on the dense network of ground motion stations deployed by AFAD revealed that the rupture that initiated on the Narli fault transitioned to supershear speeds prior to eventually triggering the EAF[23]. This initial rupture propagated along the splay fault at sub-Rayleigh speeds for 19.5 km prior to transitioning to a supershear event for the remaining length of the Narli fault before reaching the EAF[23]. Supershear ruptures generate high intensity, and largely unattenuated shock waves[24], and consequently are more efficient in dynamic triggering[25]. Furthermore, the stress field generated by a propagating supershear rupture is inherently different from that of a sub-Rayleigh rupture and is thus likely to have influenced its migration to the EAF.

The propagation speed of the rupture along the EAF is currently being debated with competing views. On one hand, through joint kinematic inversion of HR-GNSS and the ground motion data, Melgar et al.[26] suggested that the most likely estimate of the rupture speed on the EAF is 3.2 km/s for the $M_w$7.8 earthquake. This conclusion is based on an average propagation speed during the entire event sequence which is most unlikely to be representative of such a complex fault network with multiple kinks and branches which result in unsteady, and intermittent rupture propagation[4,27]. On the other hand, Okuwaki et al.[28] using potency-density tensor inversion suggests that the rupture propagation for $M_w$7.8 earthquake shows signatures of supershear propagation along EAF. Through incorporating additional stations, and emphasizing strong ground motion records near the fault within their kinematic inversions several other studies have also suggested the possibility of supershear transients[29,30].

It is quite clear from these contradicting conclusions regarding the rupture propagation speed, that we require additional insight from the mechanics that governs rupture propagation. This will complement traditional methodologies that are informed exclusively based on kinematic inversions and global fits of ground motion records. This along with abnormally high ground velocities and acceleration in near fault records near Antakya (G. Gazetas, personal communication, February 20, 2023), prompted us to scrutinize the ground motion records for characteristic signatures associated with rupture speeds, and to examine the plausibility of supershear transients beyond those observed at the triggering of $M_w$ 7.8 earthquake[23]. Furthermore, we present a detailed workflow that highlights how we can utilize mechanistic constraints on rupture propagation speeds to infer frictional properties and construct a dynamic rupture model.

To that end, we first utilize the dense seismic network provided by AFAD to study the ground motion records of stations located in near proximity of the fault trace. Through these ground motion records, we identify locations along the fault that show characteristic signatures associated with the passage of Mach Cones, which is a defining feature of supershear ruptures. We then build a 2D dynamic rupture model of the Kahramanmaraş/Pazarcik earthquake based on constraints from the ground motion records[2], field studies of the tectonic setting[6,14], and geometric features of the fault trace[17]. Through this two-fold approach we provide physical arguments to better constrain the rupture velocity profile and consequently, the frictional behavior along various fault segments. These constraints are helpful in rationalizing the conflicting predictions of rupture history by kinematic inversions and provide insights into the mechanisms that contributed to such devastation and humanitarian loss.

## Results

**Station analysis.** Figure 1 is a detailed map showing the fault trace obtained from USGS[1,17]. It also includes the estimated location of the hypocenter according to USGS[1], marked by the red star, and the location of multiple seismic stations deployed by AFAD[2]. Several of these stations are located very close to the fault surface and thus provide detailed insight into the near-field characteristics of the fault rupture. For example, Rosakis et al.[23] used stations 1 and 2 near the Narli splay fault, labeled on the map with a blue (4615) and a green circle (NAR) respectively, to show that the rupture transitioned from sub-Rayleigh to supershear rupture speeds at an epicentral distance of about 19.5 km[23]. In this study, we refer to a rupture as being supershear if we identify characteristic signatures of a propagating Mach Cone, within the near fault, ground motion records. This classification does not necessarily imply that the supershear part of the rupture front has already saturated the seismogenic zone (we refer readers to Kaneko and Lapusta[31] regarding depth saturation of supershear ruptures). It does, however, imply that this supershear portion extends to sufficient depth such that stations located within a few kilometers from the fault are dominated by its signatures. This is of particular importance as the existence of a Mach Cone within that region would have strong implications on the resulting hazard for sites located close to the fault.

Similar to Rosakis et al.[23] analysis of the Narli fault, we investigate the ground motion velocity records, resolved along the fault parallel, and fault normal directions. Furthermore, we expand our analysis to include all, complete and reliable, records from near-field stations along EAF. The raw NS, EW and vertical acceleration records are obtained from (AFAD : Disaster and Emergency Management Authority) and (KOERI : Kandilli Observatory and Earthquake Research Institute) respectively (Retrieved 02/09 5:18 PST)[2,32].

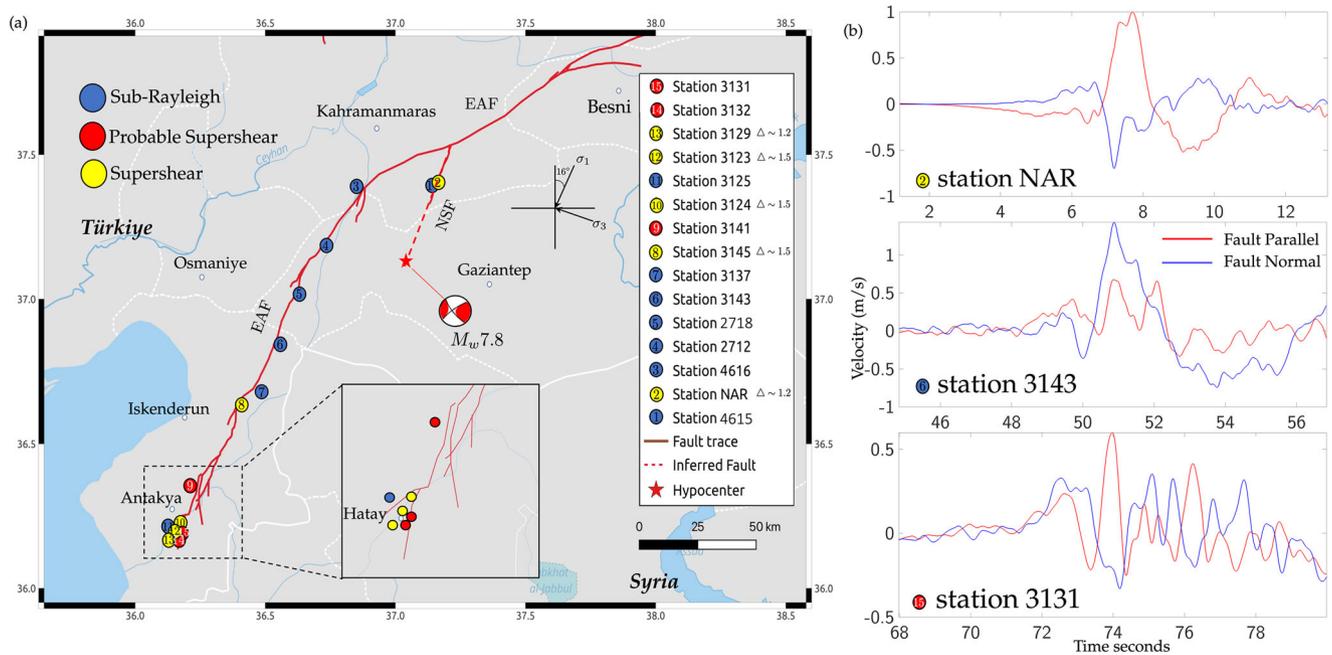

**Fig. 1 A Map of the East Anatolian Fault (EAF) highlighting the estimated location of the hypocenter of the $M_w$7.8 Kahramanmaraş/Pazarcik earthquake, and locations of major cities and towns. a** The location of seismic monitoring stations is highlighted by filled circles within the map. Stations are distinguished by their colors indicating ground record signatures consistent with either sub-Rayleigh (blue), or supershear rupture (yellow). Probable supershear is shown in red. For stations that demonstrate supershear characteristics we indicate the ratio of fault parallel to fault normal ground motion velocity components within the label. **b–e** Selected examples of the instrument-corrected ground motion records (filtered at 2 s) for stations corresponding to different rupture speed scenarios. Inserts to the figure show a zoomed view of the stations located at the southern end of the fault trace. The direction of the principal stress obtained from inversion of focal mechanism of prior earthquake history is shown on the map[6].

As discussed in Rosakis et al.[23] and Mello et al.[33], a major characteristic of supershear ruptures[34,35], which is associated with the presence of shear Mach Cones, is the existence of a dominant jump in the fault parallel ground velocity component relative to the fault normal[36,37]. Accordingly, we classify the stations based on the ratio of the fault parallel $\delta \dot{u}_{FP}^s$ to the fault normal component $\delta \dot{u}_{FN}^s$ of the velocity jump into three main categories: (1) a sub-Rayleigh station is one which experiences a dominant jump in the fault normal component, (2) a potentially supershear station is one in which the FP component is comparable to the FN component, and (3) a supershear station is one in which the FP clearly dominates the FN velocity. Similar to Mello et al.[36], we refer to a "velocity jump" as the amplitude of the maximum particle velocity swing measured from the trough to the peak of the velocity pulse associated with the passage of the rupture front at a given observation station[36]. In the legend of Fig. 1a, we provide the complete list of the stations alongside with the value of the ratio, △, of the FP to FN components of the ground velocity jumps when it represents a supershear case. This analysis allows us to identify regions along the fault where we suspect a supershear rupture has propagated during the Mw 7.8 earthquake. Figure 1b–e provides characteristic examples of the ground motion records for each rupture scenario. All the records for the other stations are included in the Supplementary Fig. 1.

The ground motion records reveal three zones in which the rupture propagation speed exceeded $C_s$. The first incident of supershear propagation, discovered in Rosakis et al.[23], occurs along the splay fault (the Narli fault) in close proximity to the hypocentral location (~20 km)[23]. After transitioning to the EAF, the rupture propagated bilaterally[3,26]. One rupture tip propagated in the NNE direction towards Malatya while the other tip propagated in the SSW direction towards Antakya. Several stations exist along the latter segment and provide sparse but important constraints on the rupture speed in that direction. Specifically, the records at stations 4 (2712), 6 (3143), and 7 (3137) show large jumps in the FN ground velocity components compared to their FP counterparts, suggesting sub-Rayleigh propagation speeds along this major segment of the EAF. Station 8 (3145) shows an opposite signature characterized by a dominant FP component jump in the ground velocity record. The ratio of the FP to FN component jumps at this station is 1.5 suggesting that the rupture is propagating at a supershear speed.

In Fig. 1, both stations 8 (3145) and 9 (3141) are located along two segments of the EAF with strikes >50° which vary substantially from the average segment strike of 25°. This suggests that the sudden change in the fault strike and the resulting change in the local stress state could have contributed to their transition to supershear speeds. Later on, near the end of the fault trace, we observe that the rupture transitioned again to supershear as indicated by the multitude of stations (10–15) located in Hatay province. Except for station 11 (3125), the other records indicate a more dominant FP to FN-component ratio. However, this ratio varies between different stations. This may be explained by the complexity of the fault network within this region. The multiple kinks and branching segments in the southern tip suggest a complex stress state that contributes to bursts of supershear on some segments and complex wave fronts that may obscure the Mach cone signature in other locations. This complexity also contributes to a stress shadowing effect on some other segments that may slow down the rupture or even prevent it from further propagation as it might have been the case for the branch near station 11 (3125).

Our analysis of the near-field station records suggests that the rupture propagation over the Narli fault as well as the SSW segment of the EAF has featured a mix of sub-Rayleigh and supershear rupture speeds. However, the sparsity of stations around the junction point of the Narli fault with the EAF, as well as along the NNE segment of the EAF, do not provide enough information to constrain the propagation speed history along that particular segment. To fill this gap, we start by developing a mechanistic model for the Narli/EAF junction consistent with the existing records on the Narli fault as outlined in the next sections.

**The Narli/EAF junction model**. To better constrain the analysis of regions with minimal station deployment, we first construct a local model focusing on the junction between the Narli fault and the EAF. This model consists of the Narli splay fault and a small portion of the EAF with the objective of obtaining better insights into the rupture migration. Figure 2a shows the region of interest and highlights the sudden change in strike at the intersection of the two faults. It further shows the simplified fault geometry in this analysis in which both fault strikes are aligned with the inferred estimates provided by USGS[1]. These approximate the actual strike based on aftershock records and the complex fault trace shown in Fig. 2a. It is important to note that in this model we consider the junction between fault A and fault B to be discontinuous. This choice is motivated by the fact that both Narli fault and EAF are young faults and highly disordered ones[14,38,39].

In our model, we adopt a linear slip-weakening friction law. Fault slip starts at a point when the shear stress $\tau$ reaches the static shear strength level, given by the product of the static friction coefficient $\mu_s$ and the fault normal compressive stress $\sigma_n$. The stress then decreases linearly with increasing slip $\delta$, over a characteristic slip-weakening distance $D_c$, to a dynamic shear strength, set by the product of the constant dynamic friction coefficient $\mu_d$ and the fault normal compressive stress $\sigma_n$.

To constrain the model, first we consider the tectonic stress state in the region. Prior studies suggest that the angle of maximum compressive stress is in a N16.4°E compression regime $(\sigma_1)$[6]. Based on this maximum horizontal stress direction, we show in Supplementary Fig. 2, that the ratio of the resolved shear stress to the normal stress on any fault segment depends on the choice of relative principal stress magnitudes. For example, using the strike of the splay fault and the orientation of the maximum compressive stress, it follows from the analysis in Supplementary Fig. 2, that any stress ratio $\sigma_1/\sigma_3$ less than 3 would result in a low apparent friction $\mu = \tau_o/\sigma_o$ ($\leq 0.3$) on the splay fault, where $\tau_o$ and $\sigma_o$ are the initial shear and normal stress respectively. That is probably inconsistent with triggering on an immature, previously unmapped, fault like the Narli fault, and it may hinder the rupture continuation on the EAF assuming reasonable values for the static and dynamic friction coefficients[40,41]. Specifically, with low apparent friction, the dynamic stress drop may be too low to enable the continued propagation past the junction. However, a stress ratio $\sigma_1/\sigma_3$ of 4 or more would increase the apparent friction to at least 0.5. This overcomes the aforementioned limitations.

Another unique constraint on the model, identified in Rosakis et al.[23], is that the rupture transitioned to supershear on the splay fault after propagating for 19.5 km at sub-Rayleigh speed. The transition to supershear depends on the frictional length scale $L_f$[42,43] and the strength parameter $S$. The strength parameter measures how close the initial stress is to the static strength $S = \frac{\mu_s - \mu}{\mu - \mu_d}$[34,44,45]. The lower $S$ value promotes a fast transition to a supershear wave, whereas the higher value indicates a favorable condition for sub-Rayleigh wave propagation[46]. Here we assume a frictional length scale $L_f = GD_c/\sigma_o(\mu_s - \mu_d) = 1600$ m ($G$ is the shear modulus), which is consistent with what is typically inferred for large crustal earthquakes[47]. We further assume that the static friction coefficient is $\mu_s = 0.7$ which is consistent with Byerlee's law[48]. To constrain the dynamic friction coefficient, we use a trial-and-error approach to obtain a value for $S$ that would yield a transition length of 19.5 km as shown in Supplementary Figure 3. We identify this value of $S$ to be 0.75. This low $S$ value is consistent with the rapid transition to supershear propagation that is inferred from near-field observation. From the known $S$ value, we then obtain the dynamic coefficient friction for the splay fault as 0.327.

Finally, given the above parameters, we adjust the value of the principal stresses to numerically produce a reasonable value of stress drop which results in a slip distribution on the splay fault that is consistent with the inferred slip from the seismic inversion (~1–3 m). We found that a minimum principal stress of $\sigma_3 = -15$ and a maximum principal stress of $\sigma_1 = -60$ MPa would produce the targeted average slip. We have also found that our choice of stress is consistent with the stress shape ratio $R$ range between 0.2–0.71 obtained from focal mechanism studies[49], where $R$ is defined as $R = \frac{S_v + \sigma_1}{\sigma_1 - \sigma_3}$, and the vertical stress $S_v$ is given as $S_v = (1 - \gamma)\rho g h$. For this range of $R$, and our choices of stress, $S_v$ at an average depth 5 km would vary between 20 and 50 MPa, which implies that $\gamma$ is between 0.6 and 0.8. We note that overpressurization of pore fluids at depth has been pointed out by earlier studies suggesting overpressurization of pore fluids at depth[50]. According to this estimate, the average slip on the splay fault is around 2.0 m and the stress drop is 3.61 MPa. Given these parameter choices, the resulting characteristic length $D_c$ corresponds to = 0.316 m. This completes the choice of parameters for the splay fault, resulting in an inferred fracture energy $G_c = \frac{1}{2}\sigma(\mu_s - \mu_d)D_c = 1$ MJ/m$^2$.

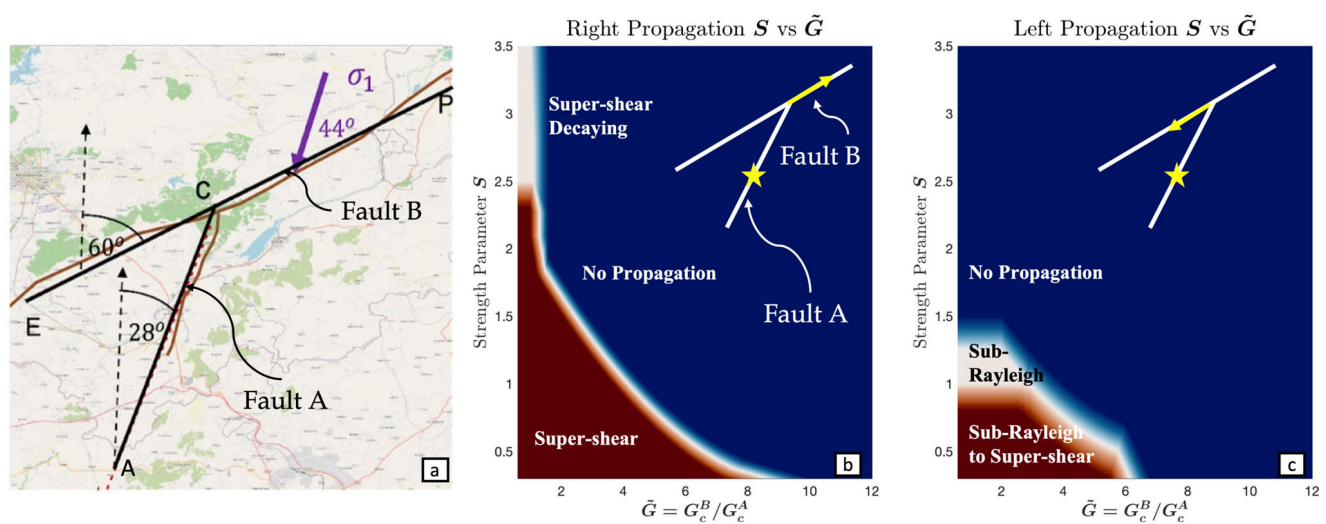

**Fig. 2 Geometry and Phase Diagram (strength parameter $S$ and ratio of fracture energies $\tilde{G}$ between splay fault (A) and main fault (B) of the Junction Model. a** The idealized geometry of splay fault (AC) and main fault (ECP) with its angle measured with respect to the North. Purple arrow represents the direction of maximum principal stress. **b** Phase diagram of right propagation (C to P direction). There are three phases: supershear propagation (brown color), supershear propagation with decaying velocity (white color), no propagation (blue color). **c** Phase diagram of left propagation (C to E direction). There are three phases: sub-Rayleigh propagation with an eventual transition to supershear (brown color), sub-Rayleigh propagation (white color), or no propagation (blue color). The critical value of $\tilde{G}$ and $S$ such that the rupture can propagate is given graphically by the boundaries of the blue region.

To investigate the implications of the constrained splay fault dynamics on the continued propagation along the EAF, we conduct a parametric study of the junction region. The objective is to constrain the frictional parameters on EAF and the properties corresponding to an early bilateral propagation beyond the junction point. To this end, we introduce a dimensionless parameter $\widetilde{G}$ which is defined as $G_c^B/G_c^A$ and correlates with the probability for continuous propagation after the jump between faults. If one considers a rupture transitioning from fault A to fault B, the parameter $\widetilde{G}$ measures the relative value of the fracture energy of fault B to the fracture energy of fault A. This quantity depends on the frictional parameters and the normal stress resolved along each individual fault. Theoretically, a small value of the $\widetilde{G}$ suggests a favorable continuous propagation due to comparable fracture energy between fault A and fault B while a large value of the $\widetilde{G}$ suggests unfavorable continued propagation. In the context of the junction, all the parameters for the splay fault (fault A) are known quantities and have been constrained using the above procedure. The objective here is to investigate the space of $S$ and $\widetilde{G}$ parameters for fault B (Line ECP) that would affect both right propagation (From C to P) and left propagation (From C to E) of the rupture on the EAF (fault B).

To conduct this investigation, we perform multiple numerical simulations modeling the rupture transition from fault A to fault B covering a wide spectrum of frictional parameters. Each individual simulation corresponds to specific choice $\widetilde{G}$ and $S$ on the EAF. In order to adjust the values of $\widetilde{G}$ and $S$ on the EAF we chose to vary frictional paramete rs $D_c$, $\mu_s$ and $\mu_d$, while keeping the $L_f$ and the normal stress fixed. The change in frictional behavior is consistent with field studies highlighting varying rock types along the EAFZ[51]. Additionally, we expect that dynamic weakening may be triggered by different mechanisms in different segments of the fault. Such mechanisms for example bimaterial contrast[52], or thermal pressurization due to different hydraulic properties would justify varying the dynamic coefficient of friction. In each of these simulations the incoming rupture on fault A was considered to be supershear as consistent with our previous discussion. Figure 2b shows the phase plot for the forward propagating front for a wide range of $\widetilde{G}$ and $S$ values. We notice that for every value of $S$ there is a critical value of $\widetilde{G}$ such that there is no propagation to the right of the junction. The relationship between that critical value of $\widetilde{G}$ and $S$ is given graphically by the boundary between the blue and the white/brown regions. We observe that as $S$ decreases the critical value of $\widetilde{G}$ required for continuous propagation increases. This can be intuitively understood as a competition between required fracture energy and fault strength. Specifically, as the fracture energy increases, the initial traction needs to be closer to the static strength to allow for continuous propagation. However, for values of $\widetilde{G}$ that permits the continued propagation, we observe that the rupture propagates as a sustained supershear, if $S$ is small enough (brown region), and as a decaying supershear if $S$ is sufficiently large ($S > 2.5$) (white region). It is obvious from Fig. 2b that if there is any rupture propagation to the right then this rupture has to initiate as a supershear rupture regardless of the choice of the parameters. This is consistent with the experimental analysis conducted by Rousseau and Rosakis 2003 which investigated the rupture propagation speed for a crack encountering a branch[53]. The study of Rousseau and Rosakis evaluated a wide spectrum of branch angles and showed that for acute branching angles (similar to the angle between the splay fault and EAF) the crack speed along the branch would initially be the same or slightly smaller than its propagation speed prior to encountering the branch[53,54].

Figure 2c shows the characteristics of the left propagating rupture in terms of the $\widetilde{G}$ and $S$ parameters. We observe that should $S > 1.5$, regardless of the $\widetilde{G}$ parameter, no back propagation will be observed. We note that $S > 1.5$ would still allow propagation to the right should $\widetilde{G}$ be small enough. Inversely, if $S < 0.9$ the rupture will back propagate initially as sub-Rayleigh prior to transitioning to supershear with the critical value of $\widetilde{G}$ increasing as $S$ decreases. For intermediate choices of $S$ ($0.9 < S < 1.5$), if $\widetilde{G}$ is small enough, the rupture can back propagate at sub-Rayleigh speeds or not propagate in the backward direction for higher values of $\widetilde{G}$. Seismic inversions reveal that there is indeed a backward propagating rupture. To further reconcile the findings for both the right and left propagation, and assuming that the frictional properties on both segments are the same, we may conclude that $S < 1.5$ and a small enough $\widetilde{G}$, would satisfy both conditions of backward propagation and sustained supershear rupture for the forward propagation.

Within the limitations of our linear elastic model, the parametric study above reveals several important findings which we summarize as follows. (1) The continuous propagation of the rupture to the right is conditional on a critical value of $\widetilde{G}$ which depends on $S$. (2) Should the supershear rupture successfully jump from the splay fault to the main fault, the rupture propagation to the right must start as a supershear. (3) The continued propagation to the right of the junction is a necessary but not a sufficient condition for the triggering of the rupture propagation to the left. This back propagating rupture additionally requires a relatively low $S$ value ($S < 1.5$). (4) If $S$ is too low ($S < 0.9$), the back propagating rupture could eventually transition into supershear. This highlights the critical dynamics of the junction and the strong dependence of the details of the rupture propagation on the stress and frictional parameters.

**2D dynamic rupture model**. After constraining the conditions that allow the bilateral propagation of the rupture on the EAF following its migration from the Narli fault, our next step is to characterize the rupture propagation along the multiple major fault segments. To that end, we consider a 2D model of a non-planar branching fault network of strike-slip faults utilizing the estimated fault trace provided by USGS based on fault offsets[1,17]. We start by generating a smooth version of the fault trace by adopting the estimated strikes of the three major segments from the USGS finite fault model for the M7.8 Kahramanmaraş/Pazarcik earthquake[1,17]. We then enrich the model at specific locations by incorporating confirmed branches and kinks. As shown in Fig. 3 the fault model consists of three primary segments spanning the two strike-slip faults: the first segment, AC, represents the Narli fault (the splay fault that hosted the hypocenter and the initial rupture propagation). The second and third segments, segments CE and CD, are both part of the EAF with different overall strike angles consistent with the USGS model[1,17]. We extend our model to capture the complexity in the fault network within the southern part between nodes F and R by incorporating multiple branches and changes in the strike. Furthermore, since the EAF is a relatively young fault and is a highly disordered one[14,38,39], we assume the fault segments are discontinuous at the locations of different geometric complexities, such as kinks and junctions between different intersecting faults. We highlight these locations with blue-filled dots in Fig. 3. Introduction of this strong segmentation may lead to transient rupture propagation interruption. However, this would still be consistent with what is expected on a geometrically complex fault system with multiple kinks, branches, and changes in strike as the one studied here.

With the frictional parameters constrained on the splay fault at hand, together with the findings after conducting the $\widetilde{G}$-$S$ parametric study in the previous section, we proceed to construct the frictional parameters appropriate for the other fault segments as follows: First, we assume that the static friction coefficient is constant for all fault segments and we set it to be $\mu_s = 0.7$. This choice is within the reasonable range for the static friction coefficients according to Byerlee's law[48]. As the rupture jumps onto the main fault (Line ED), we choose $S = 1.38$ and $\widetilde{G} = 0.755$ so that we can ensure bilateral propagation beyond the junction point C. This choice of the $S$ parameter allows supershear rupture to the north east (right) and sub-Rayleigh rupture, to the south west (left). Given an apparent friction $\mu = 0.612$ inferred from the projection of tectonic stress, this choice of S sets the dynamic friction to $\mu_d = 0.55$, and similarly this choice of $\widetilde{G}$ sets $D_c$ to be 0.275. The lower value of $\widetilde{G}$ promotes the continuous bilateral propagation along the main fault. For the fault beyond the left kink (Line EF), $S$ is assumed to be 1.621 so that sub-Rayleigh rupture is more favorable, which agrees with the signals received by the near-field stations (Fig. 1 Stations 3, 4, and 5). As for the dynamic friction parameter, all faults beyond the left kink (Point E) have a dynamic friction coefficient of 0.26. This ensures that $\mu_d < \mu$ so the dynamic propagation is facilitated by a positive dynamic stress drop. It also ensures that the parameter $\widetilde{G}$ is low enough to make it possible for the rupture to navigate the changes in strike and potentially trigger the branched segments in the southern region. Due to their orientation with respect to the background stress field, the faults located in the south end are highly stressed. With the choice of the frictional parameters outlined above, these faults ended up having small S values (~0.4) which makes supershear propagation likely, if propagation occurs along these segments. The full distribution of frictional parameters across the different fault segments is provided in Supplementary Table 1.

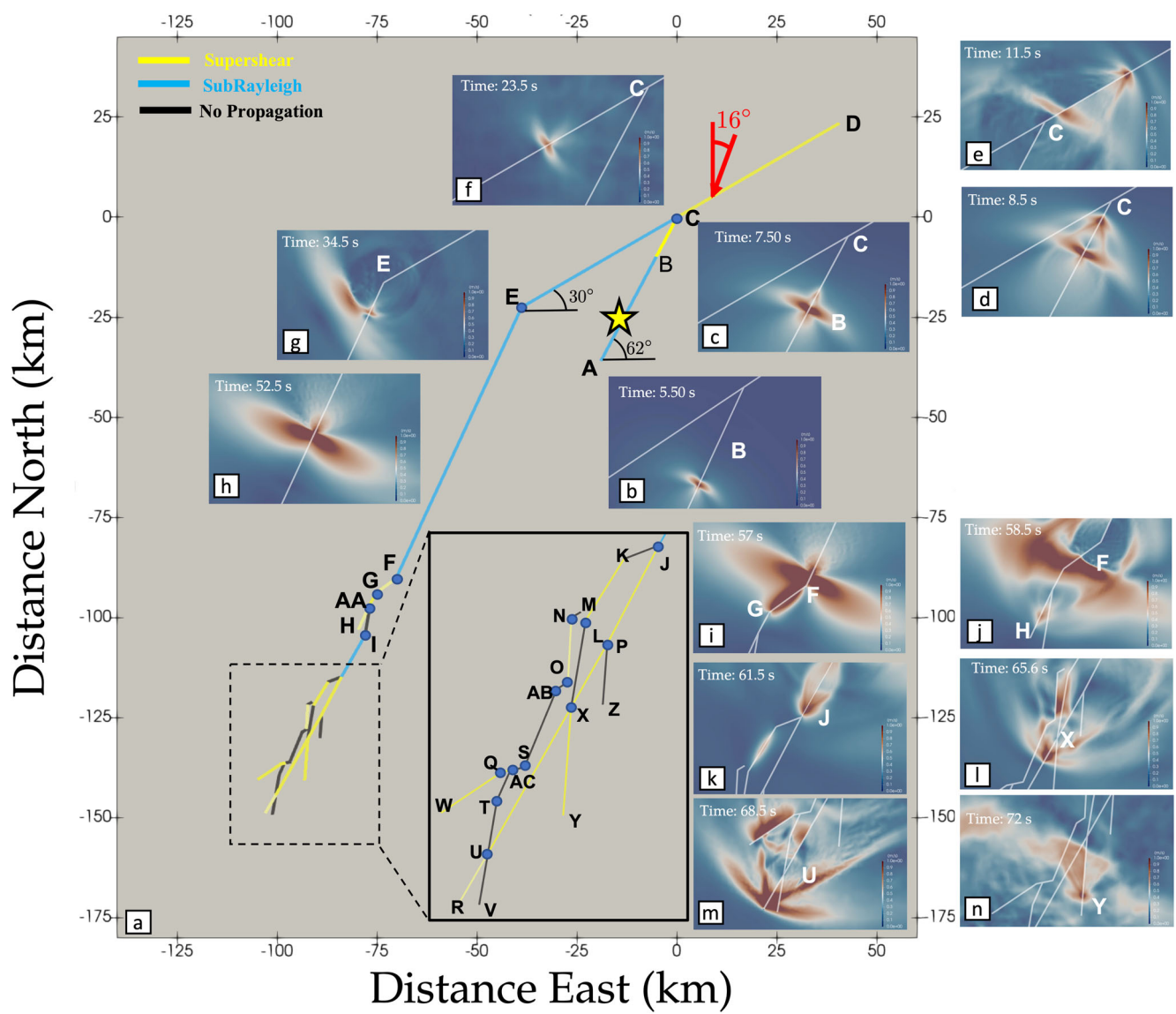

**Fig. 3 Idealized fault geometry and velocity magnitude Snapshots at specific locations along the rupture path.** Red arrow represents the direction of maximum principal stress $\sigma_1$, the yellow star is the location of the epicenter. Along the fault trace, each junction point is labeled alphabetically, while the blue dots indicate the discontinuity. Segment angles associated with junctions C and E are shown as inserts. Yellow color, blue color and black represent fault segments showing supershear, sub-Rayleigh and no rupture propagation respectively. A continuous trace of the rupture propagation speed is shown in Supplementary Figure 4b.

Figure 3 illustrates velocity magnitude snapshots of the rupture propagation at different time steps alongside a sketch of the fault system (We refer the reader to Supplementary Movie 1 and 2 for full rupture propagation history). The figure also shows the direction of the maximum horizontal principal stress, alphabetical labeling of the points of interest, a sketch of the angles for segment ED and segment AC, and blue dot marks indicating discontinuous junction points. We have also assigned different colors to mark different fault segments according to their rupture propagation speeds as will become apparent from the subsequent discussion. A continuous version of the rupture speed is also provided in Supplementary Figure 4b. The rupture is first nucleated by overstressing on the splay fault (Segment AC) with the epicenter ~30 km from the junction (Point C). The initial rupture propagates bilaterally with sub-Rayleigh speed, The rupture tip heading south arrests at the end of the splay fault (Point A). The rupture heading toward the EAF transitions to supershear speed after ~20 km of sub-Rayleigh propagation on the Narli fault (Point B, Fig. 3c). The supershear nature of the transitioned rupture is confirmed by the near-field stations (NAR), and is reproduced here with the clearly visible Mach cone in (Fig. 3c-d). As the rupture jumps onto the main fault (Line ED, Fig. 3d), the rupture to the north east (right) continues with the supershear speed (Fig. 3e) and eventually jumps into the kink point (Point D) (Line CD, Fig. 3e).

A delayed rupture to the south west (left) initiates at the junction Point C at around ~20 s from rupture nucleation, and propagates along segment CE. This time roughly agrees with the inferences based on seismic inversions[26]. This left going rupture propagates with sub-Rayleigh speed (along CE) (Fig. 3f) and jumping over the left kink (Point E, Fig. 3g). The sub-Rayleigh rupture continues propagating with increased intensity along the straight EF segment towards point F, until it reaches the region of increased geometrical complexity at the south end of the EAF.

As the sub-Rayleigh rupture approaches the end of the fault segment EF it remotely triggers a supershear rupture near Point G due to the wave field associated with incoming rupture. This supershear propagates backwards along segment GF towards Point F and merges with the incoming sub-Rayleigh rupture,(See Fig. 3i–j). This surprising propagation pattern, which is captured by the model agrees with the adjacent near field records showing that the station close to Point G (Supplementary Fig. 1h) receives the rupture signal ~0.5 s earlier than the station close to Point F (Supplementary Fig. 1g). At the same time, the same rupture propagates at supershear speed along branch GH prior to arresting at H (See Fig. 3j). As the radiated waves from the arrested phase propagates towards the southern end, a new rupture is remotely triggered along segment IP near point I by the dynamic stress field. Instead of continuous propagation, the supershear rupture gets frustrated after hitting the junctions at Point F and X. This behavior is consistent with experimental observations on interaction between rupture propagation and short pre-existing branches[55]. This rupture rapidly transitions to supershear as it continues to travel along the main fault segment IR while simultaneously activating supershear ruptures along the neighboring branches (for example Point X, Fig. 3l). This main rupture continues to propagate as supershear until it reaches the end of the fault at Point R (Fig. 3m). As shown in Fig. 1, there is a cluster of stations in this region that receives supershear signals. The fortuitous existence of a cluster of stations near the end of the fault trace, many of which record the characteristic signatures of supershear propagation, verifies the model predictions of supershear propagation near Hatay.

Our dynamic rupture model captures the following key features of the $M_w$7.8 complex event. (1) The initial nucleation of the rupture along the Narli fault and its transition to supershear at ~19.5 km away from the hypocenter. (2) The subsequent triggering of the EAF by the incoming supershear rupture. (3) The bilateral (NNE and SSW) propagation along EAF with a mix of sub-Rayleigh and supershear speeds. (4) A long portion of sub-Rayleigh growth along a major SSE segment of the EAF. (5) The supershear growth and eventual arrest of the rupture at the southernmost end of the fault trace near Hatay. Finally, the model shows that the geometric complexity and the highly heterogeneous stress field contributed to this mix of rupture speeds along different segments, as well as, additional bursts of supershear propagation along the various branches of the EAF, most notable toward Hatay.

Despite the limitation of being a 2-D model our analysis is consistent with and captures, to first order, several features associated with the strong ground motion records. In Supplementary Fig. 4a we show the ratio of the fault parallel to the fault normal particle velocity jump obtained from our numerical model. We demonstrate that the dynamic model captures the enhancement of the FP component due to supershear propagation within the regions highlighted by our station analysis as shown in Supplementary Figure 5. Furthermore, in Supplementary Figure 6 we show good agreement between synthetic arrival times obtained from our dynamic rupture model and the arrival times obtained from the ground motion records. This comparison highlights how accounting for the variable rupture speeds deduced by our station analysis to constrain dynamic rupture models can help reproduce features from the Mw7.8 Turkiye earthquake.

We further show in Supplementary Fig. 4b that the spatial distribution of rupture speed along the fault strongly correlates with the ratio of the maximum fault parallel to maximum fault normal particle velocity. Indeed we show in Supplementary Fig. 4a, b that whenever the rupture speed exceeds $C_s$, the ratio of the fault parallel to fault normal component velocity jump, $\Delta$, exceeds 1. Furthermore, as we show in Supplementary Fig. 5 the ratio of the maximum fault parallel to maximum fault normal particle velocity also increase as the rupture speed increases in the supershear regime since the FP component is increasingly enhanced by the passage of a Mach Cone.

Figure 4 a shows peak ground velocity contours for the duration of the simulated earthquake event obtained from the dynamic rupture model. Additionally, we include the distribution of the peak ground velocity in the fault parallel and fault normal direction in Supplementary Figure 5. We observe regions of intense ground velocity associated with the rupture propagation (highlighted by dashed squares). The width and extent of the intense ground motion depend on multiple factors such as the rupture propagation speed, geometrical complexity, and local frictional parameters. As highlighted earlier, the characteristics of the ground motion vary based on whether the rupture is propagating at supershear or sub-Rayleigh speeds. The intensity of the ground shaking would also depend on the stress drop which is influenced by the frictional parameters. The triggering and path selection along a complex fault network during the earthquake would play a substantial role in the distribution of PGV (peak ground velocity) within the domain. Furthermore, in the dynamic rupture model, we also observe high intensity, widely distributed ground motion near geometrical features such as the junction between the splay fault and the EAF, as well as around the left kink (Point E).

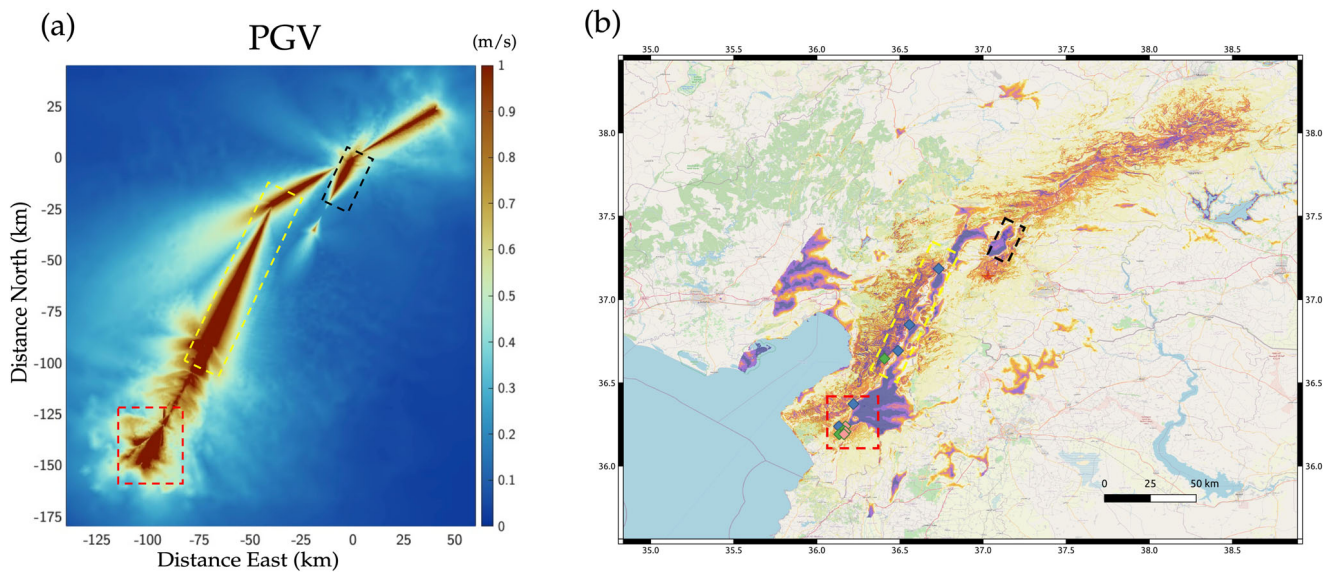

**Fig. 4 Correlation of ground shaking with ground failure estimates. a** Peak ground velocity (PGV) distribution obtained from the numerical simulation of dynamic rupture. The peak velocity distribution demonstrates regions of large magnitude PGV distribution. Geometrical complexity, triggering of segmented faults and largely unattenuated shock fronts due to supershear propagation contributes toward a wider distribution of ground shaking. **b** Ground failure estimates from USGS showing probability of liquefaction and landslide. The more extensive ground failure correlates with regions of wider and more intense ground shaking observed in our numerical model. We note that field reconnaissance of ground failure shows agreement with USGS predictions.

To associate the ground failure estimates in the $M_w7.8$ Kahramanmaraş/Pazarcik earthquake with the ground motion records obtained from the numerical model, Fig. 4b shows a map of the modeled region. On this map, we superimpose the predictions of the ground failure models generated by USGS, mainly the landslide and liquefaction estimates[1]. Both ground failure models are based on analysis of historic records of liquefaction and landslides of seismically induced ground failure. The landslide distribution models are generated based on the spatially distributed estimates of ground velocity shaking (PGV), topographic slope, lithology, land cover type, and a topographic index designed to estimate variability in soil wetness. The landslide distribution models estimated by USGS are consistent with the mapped coseismic landslides by the landslide assessment team of the 2023 Türkiye earthquake sequence (SLATE). The liquefaction model is based on slope-derived VS30, modeled water table depth, distance to coast, distance to river, distance to the closest water body, and precipitation and peak ground velocity (PGV). The liquefaction estimates from the USGS model agree with the preliminary mapping of liquefaction sites based on remote sensing data[56].

Based on both preliminary reporting and USGS estimates of ground failure we observe that regions with more distributed (mildly attenuated with distance from the fault) and intense ground motion obtained from the dynamic rupture model are consistent with regions of substantially larger destruction. The nature of the failure may be influenced by additional phenomena such as soil and basin amplification as well as the quality of construction. Supershear ruptures with intense ground motion and largely unattenuated shock fronts would probably amplify the extent and magnitude of damages associated with either structure or ground failures. Specifically, we observe that the peak slip rate rapidly changes over short distances in regions of supershear propagation to the south (Supplementary Fig. 7). This non-steady supershear propagation increases the intensity of shaking and enhances the radiated energy. Furthermore, we observe that the ground motion records show a relatively narrow (1–2 s) dominant pulse in regions with supershear propagation such as observed in Antakya (Supplementary Fig. 1lm) compared to records corresponding to sub-Rayleigh propagation (Supplementary Fig. 1d–g), a feature which was also highlighted by Wu et al.[57]. The presence of a relatively narrow velocity pulse imposes higher demand on the structures, increasing the possibility of structural collapse[58–60]. The presence of velocity pulses warrants further interrogation for destructive coherent pulses within acceleration records that can also be destructive to common civil structures[61].

Specifically, in the dynamic rupture model, we observe supershear propagation at the southern end of the fault segment in the region of Hatay near Antakya, resulting in high particle velocity magnitude (~2 m/s) and widespread ground shaking (red dashed box). Simultaneously, the records highlight substantial ground failure associated with both liquefaction and coseismic landslides within the same region. A similar pattern is also observed in NNE directions toward Malatya where we may correlate the supershear propagation in that direction with the estimates of widespread landslides in the region. Furthermore, the predicted liquefaction zone around the northern end of the Narli fault (black dashed box) also seems to correlate well with the region of supershear transition and propagation on that segment.

## Discussion
Our analysis of near-field records of the M7.8 Kahramanmaraş/Pazarcik earthquake reveals that the rupture propagation speed was spatially not uniform; rather it varied from sub-Rayleigh to supershear speeds at different sections. This is consistent with several experimental studies and numerical simulations of geometrically complex faults which demonstrated that the existence of kinks and branches may have substantial implications on the rupture terminal speed depending on the geometrical setup in relation to the orientation of the principal stresses[53–55,62]. According to the near-field records, supershear speeds are observed predominantly along the later part of the splay fault (Narli fault) that hosted the initial rupture, and at the SSW end of the fault trace within the Hatay region. Furthermore, the geometrical complexity of the fault contributed to the emergence of transient supershear ruptures as revealed by the ground motion records showing dominant fault parallel components along fault segments with steep strike changes relative to the backbone strike. Our findings reconcile the currently available seismic inversions that arrived at contradictory conclusions regarding the rupture speed.

The dynamic rupture model for the junction region between the Narli fault and the EAF allowed us to identify a regime of frictional parameters, and infer physical constraints that would be consistent with sustained propagation along both the NE and SW directions of the EAF.

We first utilize geometrical constraints together with the Narli fault station analysis of ground motion records in Rosakis et al.[23], to constrain our dynamic rupture model for the junction region between the Narli fault and the EAF. This allows us to identify a regime of frictional parameters, and to infer appropriate physical constraints that would be consistent with sustained propagation along both the NE and SW directions of the EAF. In this study, we chose to alter the frictional law parameters as opposed to introducing small-scale stress heterogeneity because (1) field evidence demonstrates varying frictional behavior across the EAFZ, (2) previous earthquake history which was limited to smaller segments of the EAF is likely to change the maturity level of the fault structure, particularly influencing its dynamic frictional behavior, (3) there are limited constraints on the local stress distribution and higher levels of uncertainty associated with slip distributions from previous earthquakes which will influence the choice of the small-scale stress heterogeneities. It is important to note that in active fault zones it is likely that both different frictional behavior and local small scale heterogeneities exist simultaneously which would influence the rupture propagation. However, there is no unique way to identify which contribution is larger.

Based on our analysis we find that sustained propagation in the NE direction of EAF necessitates that the rupture initially propagates to the north at supershear speeds. We have also found that the continued rupture propagation to the NE is necessary but not sufficient to trigger a delayed nucleation of the left propagating rupture towards the SE. The strength parameter $S$ to the SW side of the junction must also be low enough to enable the nucleation and sustainability of the left propagating rupture. Furthermore, a combination of high dynamic stress drop, on the Narli fault, and high stresses, on the EAF, appear to have been necessary to facilitate the migration of the rupture from Narli to the two sides of the EAF.

Our dynamic rupture model highlights the effect of geometrical complexity on the rupture propagation speed and the resulting complex, rupture history. Through incorporating the geometrical complexity at the intersection between the Narli fault and EAF we reproduce a major feature of this earthquake, which is the emergence of a delayed back propagation (~20 s from the rupture nucleation) to the left of the junction. Initially the angle to the left is unfavorable to sustain rupture propagation. However, the continued rupture propagation toward NNE causes stress concentration at the junction due to the dynamic stress transfer. This stress concentration eventually overcomes the static strength of the left side of the junction, which has been lowered due to tensile stress changes imparted by the incoming rupture on the Narli fault. The combination of these factors leads to a delayed nucleation and subsequent propagation in the SSW direction.

As the rupture continues to propagate in SSW direction toward Hatay, episodic supershear pulses are also seen to be triggered, in particular on small fault segments with large changes to the strike angle. This model prediction captures an interesting feature within the ground motion records. There, station 8 (Supplementary Fig. 1h) records an earlier onset of ground motion than station 7 (Supplementary Fig. 1g) which is located earlier along the fault trace. Moreover, the highly segmented nature of the EAF, which is incorporated in our model, contributes to the acceleration and deceleration of the rupture tip at different locations. In addition, this segmentation facilitates dynamic triggering, and enhances the complexity and intensity of the predicted wavefields. For example, as the rupture tip arrives just at station 8 it encounters widespread geometrical complexity which spreads all the way to the end of the fault trace. Before entering this region, the incoming rupture was propagating at sub-Rayleigh speeds. However, as it encounters the fault branches it induces bursts of supershear ruptures in most of them rather than a single strong supershear rupture. This results in an entire region dominated by complex ground shaking associated with the presence of various propagating Mach Cones. Indeed, our model reveals that the main rupture tip transitioned to supershear at Point K (shown in Fig. 3), before arriving at Antakya. This observation is consistent with both the ground motion records revealing dominant FP to FN components within the southern regions, and with the extent of ground failures, and extensive damage observed within that region

Furthermore, our numerical analysis suggests that stress and frictional conditions on the fault must have been heterogeneous. This heterogeneity contributed to the continued propagation of the rupture and influenced the rupture speed. Also, several segments of the fault are highly stressed due to their orientation with respect to the tectonic stress field. This has contributed, for example, to the early supershear transition on the Narli fault, and to the bursts of supershear propagation in the south discussed above. A combination of high dynamic stress drop on the Narli fault and a critically stressed EAF also facilitated continued propagation. Had the stress field orientation been different by a few degrees, the overall size of the event could have been much smaller.

Consistent with our station analysis, and our dynamic rupture model, a few independent dynamic rupture models have recently appeared and are in good agreement with our conclusions. Wang et al.[63] conducted a 3-D dynamic rupture model of the Turkiye earthquake and supported the conclusions of Rosakis et al.[23] that suggest the Narli fault transitioned to supershear just before reaching the junction[63]. Furthermore, they also highlighted variable propagation speeds at locations consistent with what we report in this study. Another 3-D dynamic rupture model, which was proposed by Jia et al.[64] and Gabriel et al.[65], presented results (Fig. 4b and supplementary materials movie 1) that, to our interpretation, indicate the clear presence of supershear along various segments of the fault(see for example, the end of the Narli fault in Fig. 4b)[64,65]. However, the authors report that the rupture propagates as sub-Rayleigh throughout the event. We assume that their interpretation is due to the fact that the supershear segments in their model did not fully saturate the seismogenic zone. We believe that this disagreement in speed classification stems from a difference in defining what "supershear" is. According to our definition a rupture is classified as supershear if we are able to identify characteristic signatures associated with propagating

Mach Cones, a necessary condition for such speeds. We note that ruptures in real fault zones are complex and different portions of the rupture front may propagate simultaneously at supershear, and sub-Rayleigh speeds. Consequently, relying on global measures or average speeds in rupture classification may be insufficient as the presence of shock waves, even confined near the surface, carries implications on source physics, and most importantly near-fault hazard.

While previous observations indicate that supershear ruptures are more likely to occur on long fault segments with uniform high stress, on-fault and off-fault heterogeneities can contribute to the emergence of supershear bursts as observed in our dynamic rupture model[66–69]. These transient events are difficult to identify using sparse instrumentation and far-field measurements. Furthermore, the geometric complexity may lead to complex wave fields that obscure the Mach cone signature in the far-field. Additional heterogeneity in the velocity structure may also contribute to the masking of the Mach cone in the far-field and makes it harder to detect[70,71]. This may explain the observations by Meng et al.[72] who conducted a waveform correlation analysis on the SW segment and was not able to find any persistent Mach Cone signatures in the far field. Indeed, such correlations are associated with long supershear rupture propagation rather than episodic propagation, as the ones reported here[72]. Despite the fact that supershear was not highlighted in this study, it is important to note in that the authors report faster rupture velocities along the segments that show dominant FP component, However, back projections can only predict average rupture velocities rather than local variations.

Supershear ruptures have important implications on the local hazard, even if their signature is lost in the far field. This is due to a combination of factors including (1) a narrow dominant pulse which could cause amplification of shaking for longer period structures, and (2) a largely unattenuated shear mach front. Finally, when a rupture transitions from sub-Rayleigh to supershear, there still is a sub-Rayleigh signature following the leading supershear rupture. This is called the trailing Rayleigh signature and propagates at Rayleigh wave speed[24,36,33]. As a consequence, a building at a near fault location will first experience the intense shaking due to the shock waves of the leading supershear rupture front. This part of the shaking will occur very rapidly (hence the narrow velocity pulse) and is characterized by a dominant fault parallel component of the ground velocity[36]. However, soon (seconds later) after that, the building will also experience shaking, now primarily in the fault normal direction, which is associated with the passage of the trailing Rayleigh signature. This double punch effect associated with the first (leading) arrival of the shock front and then the subsequent (trailing) Rayleigh signature can have a devastating impact on the structure. The impact of supershear ruptures on ground and structural failures warrant further investigations. Furthermore, investigations on the role of supershear rupture on the back-propagation has recently been highlighted by another study. Focused on the delay time associated with the backpropagation rupture along SW segment observed in the Turkiye earthquake Ding et al. investigated different scenarios consisting of different rupture propagation speeds along the Narli fault and indeed observed that the rupture characteristics along the Narli fault greatly influence the triggering of EAF, as well as, the delay time of backward propagation[73].

The role of physics-based dynamic modeling, especially when augmented with near-fault observations as is in the present study, is crucial to our understanding of the operant mechanism leading to such a devastating outcome. While we cannot at the current time predict the occurrences of earthquakes ahead of time, we may utilize our interpretations to better guide the response during future earthquakes.

## Methods

All numerical simulations were run using an in-house partial differential equation solver built on MOOSE framework[74]. Specifically, we utilize the cohesive zone model capability offered in TensorMechanics system[75] and implement within it a linear slip weakening law[76] as a traction-separation relation that governs the evolution of the dynamic rupture. This nonlinear solver discretizes the governing equations spatially using the finite element method and temporally using explicit time integration via the central difference method.

## Data availability

The mapped surface rupture data are from https://doi.org/10.5066/P985I7U2[17]. All the ground motion records used in this study are obtained from AFAD[2]. Fig. 1 was produced using QGIS based on map data from Natural Earth.

## Code availability

The software used to conduct the dynamic rupture model and input files required to reproduce the results presented in this manuscript is available on github (https://github.com/chunhuizhao478/farmscode.git).

## Acknowledgements

The authors would like to thank associate editor Carolina Ortiz Guerrero, Eric Dunham and an anonymous reviewer for their constructive comments that improved this manuscript. The ground motion data used in this study can be obtained from Turkish Disaster and Emergency Managment Authority AFAD, US Geological Survey (USGS), and Kandilli Observatory And Earthquake Research Institute. We would like to thank the Turkish Disaster and Emergency Management Presidency (AFAD) for setting up dense near-fault observatories, and for immediately publishing a huge number of openly accessible accelerometers during these trying times for Türkiye. A.J.R. acknowledges support by the Caltech/MCE Big Ideas Fund (BIF), as well as the Caltech Terrestrial Hazard Observation and Reporting Center (THOR). He would also like to acknowledge the support of NSF (Grant EAR-2045285). A.E. acknowledges support from the Southern California Earthquake Center through a collaborative agreement between NSF. Grant Number: EAR0529922 and USGS. Grant Number: 07HQAG0008 and the National Science Foundation CAREER award No. 1753249 for modeling complex fault zone structures. We are grateful to Idaho National Lab for providing High performance computing support and access and for the MOOSE/Falcon team for offering technical support. Funding provided by DOE EERE Geo-thermal Technologies Office to Utah FORGE and the University of Utah under Project DE-EE0007080 Enhanced Geothermal System Concept Testing and Development at the Milford City, Utah Frontier Observatory for Research in Geothermal Energy (Utah FORGE) site.

## Author contributions

M.A.: led the study, performed the station analysis, participated in the dynamic rupture modeling; designed the figures and tables; participated in the interpretation of the results; wrote the original draft. C.Z.: performed the dynamic rupture modeling; designed the figures and tables; participated in the interpretation of the results; wrote the original draft. E.Y.: performed seismic station analysis; participated in the interpretation of the results; wrote the original draft. G.G.: performed seismic station analysis; participated in the interpretation of the results; wrote the original draft. A.E. & A. R.: concieved and supervised the study and wrote original draft. All authors contributed to finalize the manuscript.

## Competing interests
The authors declare no competing interests.

## Additional information
**Supplementary information** The online version contains supplementary material available at https://doi.org/10.1038/s43247-023-01131-7.

**Correspondence** and requests for materials should be addressed to Mohamed Abdelmeguid, Ahmed Elbanna or Ares Rosakis.

**Peer review information** *Communications Earth & Environment* thanks Eric Dunham, and the other, anonymous, reviewer(s) for their contribution to the peer review of this work. Primary Handling Editors: Joe Aslin and Carolina Ortiz Guerrero. A peer review file is available.

**Reprints and permission information** is available at http://www.nature.com/reprints

**Publisher's note** Springer Nature remains neutral with regard to jurisdictional claims in published maps and institutional affiliations.